\documentstyle[prd,aps,epsf]{revtex}

\newcommand{\pabar}{\not{\!\partial}}
\newcommand{\Od}{{\cal O}}

\newcommand{\Dbar}{\not{\!{\!D}}}

\begin{document}
\draft

\input epsf \renewcommand{\topfraction}{0.8}
\twocolumn[\hsize\textwidth\columnwidth\hsize\csname
@twocolumnfalse\endcsname

\title{Production of spin $3/2$ particles from vacuum fluctuations}
\author{Antonio L. Maroto  }
\address{Astronomy Centre, 
               University of Sussex,
               Falmer, 
               Brighton,
               BN1 9QJ, U.K.}
\author{Anupam Mazumdar}
\address{Astrophysics Group, Blackett Laboratory, Imperial College London,
 SW7 2BZ,~U.K.}
\date{\today}
\maketitle
\begin{abstract}
We study the production of spin $3/2$ particles in homogeneous scalar
and gravitational backgrounds using the mode-mixing Bogolyubov method. 
Considering only the helicity $\pm 3/2$ states, we can reduce the problem
to a standard Dirac fermion calculation and apply the standard techniques
in a straightforward way. As an example we consider a specific supergravity
inflationary model and calculate the spectrum of gravitinos created 
during preheating at the end of inflation.
\end{abstract}
\pacs{PACS numbers: 98.80.Cq \hspace*{1.5cm} Imperial preprint IMPERIAL-AST
99/4-3}

\vskip2pc]

The quantization of fields in the presence of external classical
backgrounds leads to  interesting phenomena such as the production
of particles via the amplification of vacuum
fluctuations. This effect has been mainly  studied in
bosonic models, for example production of scalars or gravitons in scalar
or gravitational backgrounds. 
In addition,  this mechanism for the creation of particles is believed to be
responsible for the generation of most of the  particles that constitute the
present universe \cite{Linde}, and in fact it plays a key role in the
modern theories of preheating after inflation. In those models, the energy
of the inflaton field is resonantly converted into particles during
the period of coherent oscillations after inflation. 
This  parametric
resonance phenomenon makes the occupation number of the 
newly created bosons  grow exponentially fast and causes their spectra to be 
characterized by resonance bands. 
Recently, the resonant generation of spin $1/2$ particles  
has also  been considered in the literature 
\cite{Baacke}. In these works, it has been shown that the limit on the 
occupation number imposed by Pauli
exclusion principle is saturated and thus the non-perturbative results
deviate considerably from what is expected in a  perturbative approach.

In this work we are interested in the creation of spin $3/2$ particles 
through  the amplification of vacuum fluctuations. The generation
of such  particles in the early universe has traditionally been treated
by considering the perturbative decay of other particles \cite{Ellis,Moroi}, 
but not using the non-perturbative
approach based on the Bogolyubov transformations technique.
Some estimations of the gravitino production during inflation, based on the 
analogy with Dirac fermions can be found in \cite{Lyth}.
The spin $1/2$ case suggests that both approaches can give rise to
quite different results. 
This could be of the utmost importance
in the so-called gravitino problem: 
in supergravity models, the superpartner of the graviton field
(gravitino) is described by a spin $3/2$ particle. If such particles
are created after inflation by some mechanism (particle collision, 
vacuum fluctuations)
they could  disrupt primordial nucleosynthesis if they do not decay
fast enough, or  if they are stable particles and their masses are 
high, they could  overclose the universe. In the perturbative
approach, these facts impose stringent
constraints on both the reheating temperature and the gravitino mass
\cite{Sarkarrep}.

The calculation of spin $3/2$ particle production from vacuum fluctuations
is plagued with  consistency problems that hamper the  quantization of 
such fields
in the presence of external backgrounds. It has been known for a long
time \cite{Velo}, that a spin $3/2$ particle in
scalar, electromagnetic or gravitational backgrounds can give rise,
apart from algebraic inconsistencies, to
faster than light propagation modes. This fact completely prevents 
a consistent quantization in such cases \cite{johnson}.
The only theory in which these problems seem to be absent is supergravity,
provided the background fields satisfy the corresponding equations of motion
\cite{SUGRA}.
However, the complicated form of the Rarita-Schwinger equation
makes it very difficult to extract explicit results even in
simple backgrounds. In this paper we will 
show that when we consider  helicity $\pm 3/2$
states (which dominate the high-energy interactions of 
gravitinos \cite{Ellis,Moroi}) 
propagating in arbitrary homogeneous (and isotropic) scalar or 
gravitational backgrounds, the equations can be reduced to a Dirac-like 
equation. The 
quantization can be done along the same
lines as for Dirac spinors and therefore the standard Bogolyubov
technique \cite{Birrell} can be used to calculate the particle production.
We will also show explicitly, within a previously considered supergravity 
inflationary model, 
that the expected  amplification does take place.

The massive spin $3/2$ dynamics in flat space-time is described by the 
Rarita-Schwinger equation. We will include the scalar field 
coupling by modifying 
the mass term, (following the notation in \cite{Moroi}):
\begin{eqnarray}
\epsilon^{\mu\nu\rho\sigma}\gamma_5\gamma_\nu\partial_\rho\psi_\sigma+
\frac{1}{2}(m_{3/2}-\Phi)[\gamma^\mu,\gamma^\nu]\psi_\nu=0.
\end{eqnarray}
As usual in supergravity models we will consider Majorana spinors
satisfying $\psi_\mu=C\bar\psi_\mu^T$ with $C=i\gamma^2\gamma^0$ the
charge conjugation matrix. Contracting this equation with $\partial_\mu$
and $\gamma_\lambda\gamma_\mu$ we get:
\begin{eqnarray}
&-&
\pabar\Phi\gamma^\nu\psi_\nu+\partial^\mu\Phi\psi_\mu\nonumber \\
&+&
\frac{1}{2}(m_{3/2}-\Phi)(\pabar\gamma^\nu\psi_\nu-\gamma^\nu\pabar\psi_\nu)=0,
\label{primera}
\end{eqnarray}
and
\begin{eqnarray}
2i(\partial_\lambda\gamma^\sigma\psi_\sigma-\pabar\psi_\lambda)
+(m_{3/2}-\Phi)(\gamma_\lambda\gamma^\nu\psi_\nu+2\psi_\lambda)=0.
\label{segunda}
\end{eqnarray}
Finally contracting this last equation with $\gamma^\lambda$ we
get:
\begin{eqnarray}
i(\pabar\gamma^\sigma\psi_\sigma-\gamma^\lambda\pabar\psi_\lambda)+
3(m_{3/2}-\Phi)\gamma^\mu\psi_\mu=0.
\label{tercera}
\end{eqnarray}
When $\Phi=0$ the three equations (\ref{primera}), (\ref{segunda}) and
(\ref{tercera}) can be written as the Dirac equation plus two constraints,
i.e:
\begin{eqnarray}
(i\pabar -m_{3/2})\psi_\mu=0,\\
\gamma^\mu\psi_\mu=0,\\
\partial^\mu\psi_\mu=0.
\end{eqnarray} 
The general solution of these equations can be expanded in helicity 
$l=s/2+m$ modes:
\begin{eqnarray}
\psi^{pl}_\mu(x)=e^{-ipx}\sum_{s,m} J_{sm}u(\vec p,s)\epsilon_\mu(\vec p,m),
\label{planewaves}
\end{eqnarray}
with $J_{sm}$ the Clebsch-Gordan coefficients whose values are: 
$J_{-1-1}=J_{11}=1$, $J_{-11}=J_{1-1}=1/\sqrt{3}$ and 
$J_{-10}=J_{10}=\sqrt{2/3}$. Here $u(\vec p,s)$ are spinors with definite 
helicity $s=\pm 1$ and normalized as $u^\dagger(\vec p,r)u(\vec p,s)=
\delta_{rs}$. If we set $p^{\mu}=(\omega, p\sin \theta \cos \phi,
p\sin \theta \sin \phi, p\cos \theta)$ with $p_\mu p^\mu=m_{3/2}^2$
and $p=\vert \vec p \vert$,  then
the three spin $1$ polarization vectors are given by:
\begin{eqnarray}
\epsilon_\mu(\vec p, 1)&=& \frac{1}{\sqrt{2}} 
(0,\cos \theta \cos\phi\nonumber \\
&-&i\sin \phi,
\cos\theta \sin\phi+i\cos\phi,-\sin\theta),\\
\epsilon_\mu(\vec p, 0)&=&\frac{1}{m_{3/2}}(p,-\omega\sin \theta \cos\phi,
\nonumber\\
&-&\omega\sin\theta \sin\phi,-\omega\cos\theta),\\
\epsilon_\mu(\vec p,-1)&=&-\frac{1}{\sqrt{2}}
(0,\cos \theta \cos\phi\nonumber \\
&+&i\sin \phi,
\cos\theta \sin\phi-i\cos\phi,-\sin\theta).
\end{eqnarray}
The normalization is $\epsilon_\mu^*(\vec p,m)\epsilon^\mu(\vec p,n)
=\delta_{mn}$,
$p^\mu\epsilon_\mu(\vec p,m)=p^\mu\epsilon_\mu^*(\vec p,m)=0$. The 
corresponding quantization details can
be found elsewhere \cite{Moroi}.

Now we turn to the $\Phi\neq 0$ case. The expression in (\ref{planewaves}) 
is no longer a 
solution of the equations of motion. Let us now concentrate on homogeneous
scalar fields, only dependent on the time coordinate $\Phi(t)$. We look 
for general homogeneous solutions of the Rarita-Schwinger equation
of the form:
\begin{eqnarray}
\psi^{pl}_{\mu}(x)=e^{i\vec p \cdot \vec x}f^{pl}(t)\sum_{s,m}J_{sm}
u(\vec p, s)
\epsilon_\mu(\vec p,m)
\label{ansatz}
\end{eqnarray}
These fields satisfy the condition $\gamma^\mu\psi_\mu=0$, since they
differ from (\ref{planewaves}) in just a scalar factor. Now if we restrict
ourselves to the helicity $l=\pm 3/2$ states, they satisfy 
$\psi^{p\pm 3/2}_0=0$
and, since  the spatial derivatives of the scalar field vanish 
$(\partial_i\Phi=0)$, then (\ref{primera}) and (\ref{tercera}) are 
automatically
satisfied provided $\partial^i\psi_i=0$. From (\ref{ansatz}) this last 
condition
is equivalent to $p^i\psi_i=0$ which holds from the
condition $p^\mu\epsilon_\mu(\vec p,m)=0$. Accordingly, for helicity 
$\pm 3/2$ states propagating in an homogeneous scalar background, the 
Rarita-Schwinger equation reduces again to the Dirac form:
\begin{eqnarray}
(i\pabar -m_{3/2}+\Phi(t))\psi^{\pm 3/2}_\mu=0
\end{eqnarray}
As far as these modes satisfy a Dirac-like equation, it appears that 
all the difficulties in
the quantization  would concern just the 
helicity $\pm 1/2$ modes in this case. 
In fact the 
above ansatz (\ref{ansatz})
is not a solution for the helicity $\pm 1/2$ modes even for homogeneous 
backgrounds.

Let us include the effect of curved space-time. We will concentrate
on spatially flat Friedmann-Robertson-Walker (FRW) metrics, 
and we will introduce it
by {\it minimal} coupling as done in supergravity, i.e, 
$D_\rho \psi_\sigma=(\partial_\rho+\frac{i}{2}
\Omega^{ab}_\rho\Sigma_{ab})\psi_\sigma$ with $\Omega^{ab}_\rho$ the 
spin-connection coefficients and $\Sigma_{ab}=\frac{i}{4}[\gamma_a,\gamma_b]$.
The $\epsilon^{\mu\nu\rho\sigma}$ removes the Christoffel symbols
contribution in the covariant derivative.
We will continue considering $\Phi(t)$ to be a function of time alone.
We will only  consider  the linearized equation in $1/M$ 
(where $M_P^2=8\pi M^2$) for supergravity  
\cite{Bailin}, i.e, we will ignore the torsion contribution 
to the spin-connection which is of $\Od(M^{-2})$.
In this case the equations of motion for the gravitino read:
\begin{eqnarray}
\epsilon^{\mu\nu\rho\sigma}\gamma_5\gamma_\nu D_\rho\psi_\sigma+
\frac{1}{2}(m_{3/2}-\Phi)[\gamma^\mu,\gamma^\nu]\psi_\nu=0.
\end{eqnarray}
Contracting with $D_\mu$, taking into account that $D_\mu\gamma_\nu=0$
and  $[D_\mu,D_\rho]=-\frac{i}{2}R^{ab}_{\;\;\;\mu\rho}\Sigma_{ab}$
(the vector part cancels because of the $\epsilon^{\mu\nu\rho\sigma}$ term), 
we get:
\begin{eqnarray}
-\frac{i}{4}\epsilon^{\mu\nu\rho\sigma}\gamma_5\gamma_\nu
R^{ab}_{\;\;\;\mu\rho}\Sigma_{ab}\psi_\sigma-(\Dbar\Phi)\gamma^\nu\psi_\nu
+(D^\mu\Phi)\psi_\mu \nonumber\\
+\frac{1}{2}(m_{3/2}-\Phi)(\Dbar\gamma^\nu\psi_\nu-\gamma^\nu\Dbar\psi_\nu)=0.
\label{primera2}
\end{eqnarray} 
Following the same steps as in flat space-time we  obtain 
(\ref{segunda}) and (\ref{tercera}) but replacing ordinary derivatives
by covariant ones. For FRW metrics and  
helicity $\pm 3/2$ states, i.e., $\psi_0=0$, 
it is possible to show that due
to the form of the Riemann tensor, the first term 
in (\ref{primera2})
is proportional to $\gamma^\mu \psi_\mu$ and accordingly  we get:
\begin{eqnarray}
(i\Dbar -m_{3/2}+\Phi)\psi_\mu=0, \label{cons0}\\
\gamma^\mu \psi_\mu=0, \label{cons1}\\
D^\mu \psi_\mu=0. \label{cons2}
\end{eqnarray}
Here again we can use the standard formulae for particle production
obtained for the spin $1/2$ cases to study the creation of helicity $\pm 3/2$
states in a FRW background. With that purpose we have to reduce equation 
(\ref{cons0}) to a second order equation. Let us first write the equation in
conformal time defined as $dt=a(\eta)d\eta$:
\begin{eqnarray}
\left(ia^{-1}\delta^\mu_a\gamma^a\partial_\mu-m_{3/2}+\Phi
+i\frac{3}{2}\frac{\dot a}{a^2}\gamma^0\right)\psi_\mu=0, 
\label{dirconf}
\end{eqnarray}
where $\dot a=da/d\eta$. We will adopt the following ansatz for the
helicity  $l=\pm 3/2$ solutions:
\begin{eqnarray}
\psi_\mu^{pl}(x)=a^{-3/2}(\eta)e^{i\vec p \cdot \vec x} U_\mu^{\vec p l}(\eta),
\end{eqnarray}
with 
\begin{eqnarray}
U_\mu^{\vec p l}(\eta)&=&\frac{1}{\sqrt{\omega+m_{3/2}^0}}\left(
i\gamma^0\partial_0
-\vec p\cdot \vec\gamma\right.\nonumber \\
 &+&\left.\left. a(\eta)(m_{3/2}-\Phi(\eta)\right)\right)f_{pl}(\eta)u(\vec p, s)
\delta^a_\mu\epsilon_a(\vec p, m),
\end{eqnarray}
and the normalization 
$U_\mu^{\vec p l\dagger} (0)U^\mu_{\vec p l}(0)=2\omega$ and 
$m^0_{3/2}=a(0)m_{3/2}$.
One can check that this ansatz automatically satisfies 
(\ref{cons1}) and (\ref{cons2}). An appropriate form for
the spinor $u(\vec p, s)$ and polarization
vectors $\epsilon_a(\vec p, m)$ can be obtained if we choose 
the Dirac representation
for the gamma matrices and we take (without loss of generality) 
the $z$-axis to be along the $\vec p$ 
direction. In this case $u(\vec p,1)^T=(1,0,0,0)$, $u(\vec p, -1)^T=(0,1,0,0)$,
$\epsilon_a(\vec p, 1)=\frac{1}{\sqrt{2}}(0,1,i,0)$ and
$\epsilon_a(\vec p, -1)=\frac{1}{\sqrt{2}}(0,1,-i,0)$. With this
choice, 
$u(\vec p,\pm 1)$ are eigenstates of $\gamma^0$ with eigenvalues $+1$.
Then equation (\ref{dirconf}) reduces to the well-known form:
\begin{eqnarray}
\left(\frac{d^2}{d\eta^2}\right.&+&p^2-i\frac{d}{d\eta}\left.\left(
a(\eta)(m_{3/2}-\Phi(\eta)\right)\right) \nonumber \\
&+&\left.
a^2(\eta)(m_{3/2}-\Phi(\eta))^2\right)f_{pl}(\eta)=0.
\label{master}
\end{eqnarray}
In order to quantize the modes we will expand an arbitrary solution
with helicity $l=\pm 3/2$ as:
\begin{eqnarray}
\psi^{l}_\mu(x)&=&\int \frac{d^3p}{(2\pi)^3 2\omega}a^{-3/2}(\eta)
\left(e^{i\vec p\cdot \vec x}U_\mu^{\vec p l}(\eta)a_{\vec p l}
\right.\nonumber\\
&+&\left.
e^{-i\vec p\cdot \vec x}U_\mu^{\vec p l C}(\eta)a_{\vec p l}^\dagger\right),
\end{eqnarray}
where the creation and annhilation operators satisfy the anticommutation
relations $\{a_{\vec p l},a_{\vec p' l'}^\dagger\}
=(2\pi)^3 2\omega \delta_{ll'}\delta(\vec p-\vec p')$.
 
In order to see how this works in practice, we will consider a specific
supergravity inflationary model (see \cite{Holman,Sarkar}), in which the inflaton
field is taken as the scalar component of a chiral superfield, and its
potential is derived from the superpotential $I=(\Delta^2/M)(\phi-M)^2$. 
This is the simplest choice that satisfies 
the conditions that supersymmetry
remains unbroken in the minimum of the potential and that the present
cosmological constant is zero. The observed CMB anisotropy fixes the 
inflationary scale
around $\lambda\equiv \Delta/M\simeq 10^{-4}$. For the sake of
simplicity, we will consider the case in which the gravitino mass is
much smaller than the effective mass of the inflaton in this model, 
$m_{3/2}\ll m_{\phi}\simeq 10^{-8}M$ and since the production
will take place during a few inflaton oscillations, 
we will neglect the mass term in 
the equations.
The scalar field potential is  given by \cite{Bailin}:
\begin{eqnarray}
V(\phi)=e^{\vert \phi \vert ^2/M^2}\left(\vert \frac{\partial I}{\partial \phi}
+\frac{\phi^*I}{M^2}\vert ^2-\frac{3\vert I\vert ^2}{M^2}\right).
\end{eqnarray}
For the above superpotential, the imaginary direction  is known to be
stable 
and
therefore we will take for simplicity a real inflaton field. Along the
real direction the potential can be written as \cite{Sarkar}:
\begin{eqnarray}
V(\phi)=\lambda^4 e^{\phi^2}\left((2(\phi -1)+\phi(\phi-1)^2)^2-3(\phi-1)^4
\right)
\end{eqnarray}
where we are working in units $M=1$. We will assume, as indicated 
in \cite{Sarkar},
that the potential contributions of dilaton and moduli fields are fixed
during and after inflation.
This potential has
a minimum at $\phi=1$. The coupling of the inflaton field
to gravitinos is given by the following mass term in the supergravity
lagrangian \cite{Bailin}:
\begin{eqnarray}
{\cal L}=-\frac{1}{4}e^{G/2}\bar \psi_\mu
[\gamma^\mu,\gamma^\nu]\psi_\nu,\\
e^{G/2}=\lambda^2 e^{\phi^2/2}(\phi-1)^2,
\end{eqnarray}
where we have chosen the minimal form for the  K\"ahler potential 
$G(\Phi,\Phi^\dagger)=
\Phi^\dagger \Phi+\log \vert I \vert ^2$. The rest of interaction terms in the
supergravity lagrangian are not relevant for our purposes.
The inflaton and Friedmann equations can be
written in conformal time as:
\begin{eqnarray}
\ddot \phi +2 \frac{\dot b}{b}\dot \phi +\frac{b^2}{\lambda^4}V_{,\phi}=0,\\
\frac{\dot b^2}{b^2}=\frac{1}{3}\left(\frac{1}{2}\dot \phi^2
+\frac{b^2}{\lambda^4}V\right),
\end{eqnarray}
where the derivatives are with respect to the  new time coordinate 
$\tilde \eta=a_0\lambda^2 \eta$ and the new scale factor is defined as 
$b(\tilde \eta)=a(\tilde\eta)/a_0$ with $a_0=a(0)$.
The solution of this equation shows that after the inflationary
phase, the scalar field starts oscillating around the minimum of the
potential with damped amplitude.
Substituting in (\ref{master}) for this particular case we obtain:
\begin{eqnarray}
\left(\frac{d^2}{d\tilde\eta^2}+\kappa^2+\frac{i}{\lambda^2}
\frac{d}{d\tilde\eta}(be^{G/2})+
\frac{b^2}{\lambda^4} e^{G}\right)f_{\kappa l}(\tilde\eta)=0,
\label{master2}
\end{eqnarray}
with $\kappa=p/(a_0\lambda^2)$. From this expression we see that
when the scalar interaction is switched off, there is no particle
production, even in the expanding background.
 Following \cite{Baacke,Mostepanenko}
we can calculate the occupation number:
\begin{eqnarray}
N_{\kappa l}(\tilde T)&=&\frac{1}{4\kappa}\left(2 \kappa
+i[\dot f^*_{\kappa l}(\tilde T) 
f_{\kappa l}(\tilde T)- f^*_{\kappa l}(\tilde T)
\dot f_{\kappa l}(\tilde T)]\right.\nonumber \\
&-&\left.\frac{2}{\lambda^2}be^{G(\tilde T)/2}
\vert f_{\kappa l}(\tilde T)\vert ^2 ]\right)
\label{occupation}
\end{eqnarray}
In order for the particle number to be well defined, we must evaluate
it when the interaction is vanishingly small, that is, for large values 
of $\tilde T$. 
Here $f_{\kappa l}$ is a solution of equation (\ref{master2})
with initial conditions $f_{\kappa l}(0)=1$ and 
$\dot f_{\kappa l}(0)=-i\kappa$ which corresponds to a plane wave for
$\tilde \eta\leq 0$. In order to define the initial vacuum
 at $\tilde \eta=0$, we have taken the inflaton to be at the minimum of
the potential at that moment ($\phi(0)=1$), which implies  
$e^{G(\phi=1)/2}=0$ and $b(0)=1$. We have chosen $\dot \phi(0)=1.8$
in our numerical computations which corresponds to  an initial amplitude
of the inflaton oscillations of about $0.06M_p$
and a maximum value of the coupling
$e^{G/2}$ of $10^{-10}M_p$.
\begin{figure}
\begin{center}
\mbox{\epsfysize=7cm\epsfxsize=7cm
\epsffile{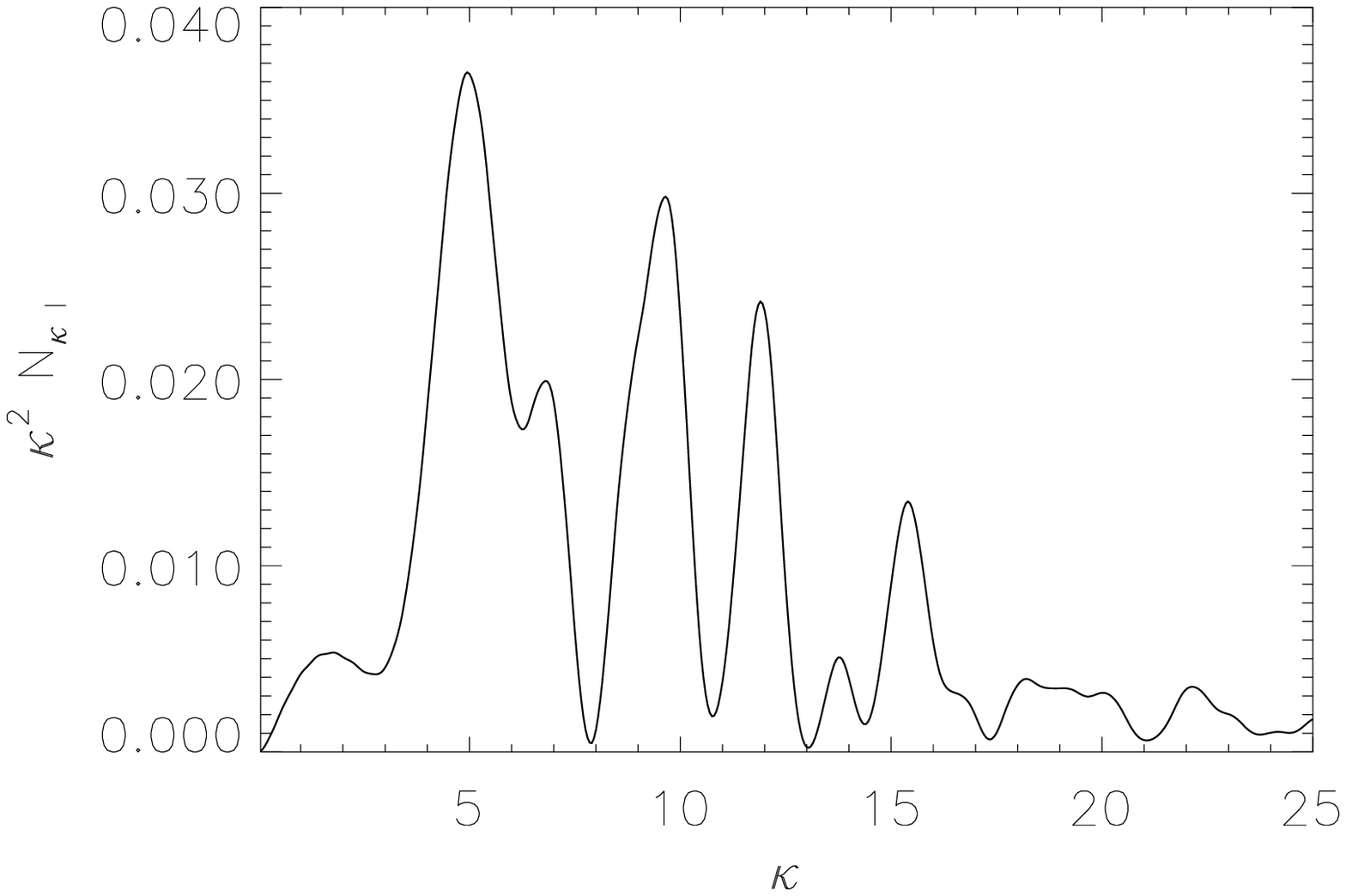}}
\end{center}
\vspace {-.7cm}
\leftskip 1cm
\rightskip 1cm
{\footnotesize
{\bf Figure 1.-}Number density of helicity $l=\pm 3/2$ gravitinos ($\kappa^2
N_{\kappa l}$)  against
$\kappa$. } 
\end{figure}
The results for the spectra in the expanding background are shown in Fig.1. 
Note that
we have not considered the backreaction effects of the produced
particles. 
In the
flat space calculation, we find  that broad resonance bands may
appear, similar
to those in  \cite{Baacke,MaMa}. When expansion 
is taken into account (Fig.1), 
the production is  reduced by 3-4 orders of
magnitude, however the number 
of particles  produced is not negligible.
From Fig. 1, we can estimate a lower bound to  the total number density of 
gravitinos of both helicities as:
\begin{eqnarray}
n(\eta)=\frac{1}{\pi^2 a^3(\eta)}\int_{0}^{\infty} N_{pl}p^2 dp=
 \frac{a_0^3\lambda^6}{\pi^2 a^3}\int_{0}^{\infty} 
N_{\kappa l}\kappa^2 d\kappa
\label{numer}
\end{eqnarray}
Comparing with
the number density of a thermal distribution of helicity $\pm 1/2$
 gravitinos as
estimated in \cite{Pagels} (the helicity
$\pm 3/2$ could be even less abundant)
for a typical
value of the scale factor at the end of inflation
\cite{Kolb} $a_0\simeq 10^{-26}$,
the vacuum fluctuation production is suppressed by a factor $ 10^{10}$. 
The corresponding cosmological consequences have been studied in  
\cite{Ellis,LiEll}. Comparing with the entropy density today we obtain:
$n/s \geq 10^{-12}$. This result is 4-5 orders of magnitude 
larger than the perturbative production during reheating
from direct inflaton decay \cite{Sarkar} and it could pose
compatibility 
problems with the 
nucleosynthesis bounds \cite{Moroi,Sarkarrep} for some values of
the gravitino mass. 

We have considered the production
of helicity $\pm 3/2$ gravitinos (which are the relevant states for the current
nucleosynthesis bounds) in a particular inflationary 
model. The expression (\ref{numer}) shows that the results are
very sensitive to the model parameters, but they can be used to
discriminate between different supergravity inflationary models. The 
completion of the picture would
require to study  other models and also include
the production of helicity $\pm 1/2$ modes;
however, the  Bogolyubov technique appears very involved for this 
purpose. (After the appearance of this work, the helicity $\pm
1/2$ case was considered in  Kallosh et al., hep-th/9907124)
{\bf Acknowledgments:} A.L.M. acknowledges support from  SEUID-Royal Society
and (CICYT-AEN97-1693)(Spain). A.M. is supported by INLAKS
and an ORS award. We thank Andrew Liddle for valuable discussions.

\vspace{-0.3cm}
\thebibliography{references} 
\vspace {-0.9cm}
\bibitem{Linde} L. Kofman, A. D. Linde and A.A. Starobinsky,
{\it Phys. Rev. Lett.} {\bf 73} (1994) 3195;
L.Kofman, A.D. Linde and A. A. Starobinsky, {\it Phys. Rev.} {\bf D 56}
(1997) 3258; J.H. Traschen, R.H. Brandenberger, {\it Phys. Rev.} {\bf D42}
(1990), 2491; Y. Shtanov, J. Traschen and R. Brandenberger, {\it Phys. Rev.}
{\bf D51}(1995), 5438
\bibitem{Baacke} J. Baacke, K. Heitmann and C. Patzold, {\it Phys. Rev.}
 {\bf D58} 125013 (1998); P.B. Greene and L. Kofman, {\it Phys. Lett.} 
{\bf B448}, 6 (1999)
\bibitem{Ellis} J. Ellis, J.E. Kim and D.V. Nanopoulos, {\it Phys. Lett.}
{\bf 145B}, 181 (1984)
\bibitem{Moroi} M. Kawasaki and T. Moroi, {\it Prog. Theor. Phys.} {\bf 93},
879 (1995); T. Moroi, PhD Thesis (1995), hep-ph/9503210 
\bibitem{Lyth} D.H. Lyth, D. Roberts and M. Smith {\it Phys. Rev.} {\bf D57}
7120 (1998)
\bibitem{Sarkarrep} S. Sarkar, {\it Rep. Prog. Phys.} {\bf 59}, 1493 (1996)
\bibitem{Velo} G. Velo and D. Zwanziger, {\it Phys. Rev} {\bf 186} (1969) 
1337; C.R. Hagen and L.P.S. Singh, {\it Phys. Rev.} {\bf D 26}
(1982) 393
\bibitem{johnson} K. Johnson and E.C.G. Sudarshan, {\it Ann. Phys.}(N.Y.)
13 (1961) 126
\bibitem{SUGRA} S. Deser and B. Zumino, {\it Phys. Rev. Lett.} {\bf 38}
(1977) 1433; {\it Phys. Lett.} {\bf 62B}, 335 (1976)
\bibitem{Birrell} N.D. Birrell and P.C.W.
Davies {\it Quantum Fields in Curved Space}, Cambridge University Press 
(1982)
\bibitem{Bailin} D. Bailin and A. Love, {\it Supersymmetric Gauge Field
Theory and String Theory}, IOP, Bristol, (1994)
\bibitem{Holman} P. Holman, P. Ramond and G.G. Ross, {\it Phys. Lett.}
{\bf 137B}, 343 (1984)
\bibitem{Sarkar} G.G. Ross and S. Sarkar, {\it Nucl. Phys.}{\bf B461}
(1996) 597
\bibitem{Mostepanenko} V.M. Mostepanenko and V.M Frolov, 
{\it Sov. J. Nucl. Phys.}
{\bf 19} 451 (1974); A.A. Grib, S.G. Mamayev and V.M. Mostepanenko {
\it Vacuum Quantum Effects in Strong Fields}, Friedmann Laboratory 
Publishing, St. Petersburg (1994)
\bibitem{MaMa} A.L. Maroto and A. Mazumdar, {\it Phys. Rev.} {\bf D59}, 
083510 (1999)
\bibitem{Pagels} H. Pagels and J. Primack, {\it Phys. Rev. Lett.} {\bf 48}
223 (1982)
\bibitem{Kolb} E.W. Kolb and M.S Turner, {\it The Early Universe} 
(Addison-Wesley) (1990)
\bibitem{LiEll} J. Ellis, A. Linde and D. Nanopoulos, {\it Phys. Lett.}
{\bf 118B}, 59 (1982)
\end{document}